\documentclass[twocolumn, DIV20, parskip = false]{scrartcl}
\usepackage{graphicx}
\usepackage{amsmath}
\usepackage{SIunits}
\usepackage{authblk}
\usepackage{cite}
\setkomafont{disposition}{\normalcolor\rmfamily}
\setkomafont{caption}{\small\mdseries\selectfont}

\setcapindent{0mm}%
\begin{document}
\twocolumn[ 
\title{\huge Embedding defect sites \\into hexagonal nondiffracting wave fields}
\author{\large Andreas Kelberer, Martin Boguslawski,$^{*}$ Patrick Rose, and Cornelia Denz}
\affil{\normalsize Institut f\"ur Angewandte Physik and Center for Nonlinear Science (CeNoS), \\Westf\"alische Wilhelms-Universit\"at M\"unster, 48149 M\"unster, Germany
\\
$^*$Corresponding author: martin.boguslawski@uni-muenster.de
}
\date{}
\maketitle
\begin{abstract}
\begin{quote}
\small We present a highly purposive technique to optically induce periodic photonic lattices enriched with a negative defect site by using a properly designed nondiffracting beam. As the interference of two or more nondiffracting beams with adequate mutual spatial frequency relations in turn reproduces a nondiffracting beam, we adeptly superpose a hexagonal and a Bessel beam to create the nondiffracting defect beam of demand. The presented wavelength-independent technique is of utmost universality in terms of structural scalability and does not make any specific requirements to the photosensitive medium. In addition, the technique is easily transferable to all pattern-forming holographic methods in general and its application is highly appropriate e.g. in the fields of particle as well as atom trapping.
\end{quote}
\end{abstract}
\bigskip
]
\noindent
Periodic dielectric media featuring structural sizes in the regime of the wavelength of a propagating wave field are notedly capable to dramatically modify the propagation behavior of light in comparison to the characteristics in a bulk material and are thus of high interest to fundamental physics as also to new technical applications \cite{Eisenberg,Fleischer,Fischer}. 
In analogy to the behavior of electrons in solid state physics, many novel phenomena of light guidance such as Bloch oscillation, Zener tunneling, or Anderson localization were firstly predicted theoretically and later demonstrated in experiments \cite{Peschel,Trompeter,Schwartz}. 
Light propagation in periodic media is described by photonic band structures where, depending on the periodical formation, eigenmodes in terms of Bloch waves can propagate on different bands, while wave propagation between the bands or rather inside so-called band gaps is forbidden \cite{Joannopoulos}. 
To make these media more functional, such a behavior can be broken by introducing artificial defects into periodic structures, causing the appearance of additive propagation modes inside the band gaps \cite{Russel,Makasyuk,WangJ1}. 
Hence, many unexplored effects have been observed such as defect solitons in waveguide arrays \cite{PeschelA}, linear vortex defect modes \cite{Song}, or nonlinear beam deflection on defect sites \cite{WangJ2}. 
However, the arbitrary implementation of largely extended two-dimensional (2D) periodic structures of high structural accuracy is still a challenging task, as common techniques of serial writing proceeding, e.g. direct laser writing techniques, reveal various disadvantages for instance in terms of writing time and a huge demand of adjustment accuracy \cite{Zeuner}.

In contrast to that, the usage of nondiffracting (ND) wave fields to write photonic lattices in a parallel manner has become a widely-approved and purposive principle \cite{Fleischer, Rose, Langner, WangL}. 
Consequently, implementing defect bearing periodic and quasiperiodic structures by a single-shot ND beam induction would be an enormous improvement in terms of accuracy and temporal effort.
Already in 2006, Makasyuk and co-authors \cite{Makasyuk} successfully implemented 2D lattices with a single-site negative defect into photorefractive crystals by utilizing a particularly balanced polarization state of the ND writing beam especially adapted to the anisotropic nonlinear behavior of strontium barium niobate (SBN) crystals to avoid self-healing effects. 
Nonetheless, this technique strongly depends on the characteristic of the used photorefractive medium. 
However, a general and medium-independent approach of parallely inducing defect enriched photonic lattices is missing up to now.

In this Letter, we propose a so far unexploited method to generate ND defect beams by coherently superimposing periodic lattices belonging to the family of discrete ND beams \cite{Boguslawski, Becker} with a certain beam of one of the four different families of ND beams \cite{curvNdbs1, curvNdbs2, curvNdbs3}.
More precisely, we combine a hexagonal lattice wave field $E_\text{H}$ with a Bessel beam $E_\text{B}$ of zeroth order to induce a local reduction of intensity at one lattice site without significant disturbance of the remaining lattice periodicity. As for all ND beams the structural size is adaptable by manipulating the respective transverse $k$-spectrum of the wave field while the general structure is maintained.

To avoid longitudinal modulation along the propagation direction of a ND beam the parallel component of all interfering wave vectors $k_\parallel$ for a monochromatic wave has to be equal, so that the value of the transversal part $k_\bot$ is also the same.
Emanating from this condition, all possible wave vectors can be arranged in Fourier space on a cone with an opening angle of $\theta$ with respect to the normal vector of the transverse plane.
The absolute value of the transverse part of the wave vector can also be determined by the equation $k_\bot = 2\pi \sin \theta /\lambda$ and leads in the case of a periodic field distribution to the lattice constant $g = \pi/k_\bot$ \cite{Boguslawski}.
To generate a superimposed light field which in addition maintains the ND properties, attention must be paid to the transverse part of the contributing wave vectors which have to be equal for both light fields $k_{\bot,\text{H}} = k_{\bot,\text{B}} = k_\bot$.
Then the superposition yielding a ND beam can be generated by the prescription of the resulting light field 
\begin{eqnarray}\label{eq:resultField}
E_\text{Res}(k_\bot) = E_\text{H}(k_\bot) + a e^{i\phi}E_\text{B}(k_\bot),
\end{eqnarray}
where the factor $a$ varies the defect strength and $\phi$ denotes a free phase parameter exploiting the phase distribution of the fields.

\begin{figure}[t]
\centering
\includegraphics[width = \columnwidth]{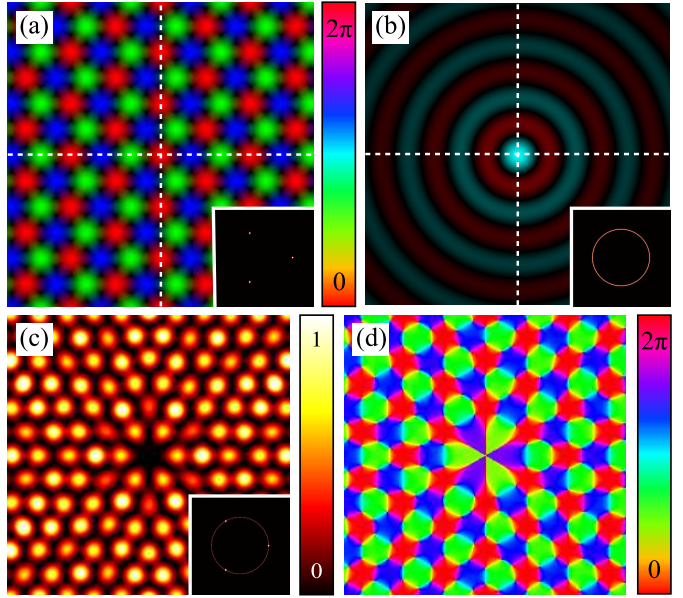}
\caption{(Color online) 
Light distribution of (a) a hexagonal beam and (b) a Bessel beam where the color represents the phase distribution and brightness is linear to varying amplitude.
Figures (c) and (d) depict the intensity and phase distribution of the combined lattice. In the insets the intensities of the Fourier spectrum are presented according to the corresponding light fields.
\label{figure1}
}
\end{figure}

The results of the numerical calculation of the discrete and Bessel beam as well as of the superimposed beam is shown in Fig. \ref{figure1}.
The hexagonal ND beam as presented in Fig. \ref{figure1}(a) results by arranging in the Fourier space three plane waves on a ring with an equidistant azimuthal shift to each other as illustrated in the inset.
For the zeroth order Bessel beam the wave vectors are distributed continuously on the ring holding mutual constant phase, so that the intensity distribution in real space has a central maximum [Fig. \ref{figure1}(b)].
The local reduction of intensity by interference of complex light fields in a certain position is effected by a phase difference of $\pi$, so that in our case the parameters for a central defect site with zeroth intensity have to be chosen as follows [cf. Eq. (\ref{eq:resultField})]: $a = 1$ and $\phi = \pi$.
The results of the superposition for intensity and phase distribution are shown in Fig.  \ref{figure1}(c) and \ref{figure1}(d), respectively.
Due to the ring shaped maxima of the Bessel beam the intensity of the lattice sites is slightly changed but still arranged periodically and the disturbance decays with increasing distance to the center.
The intensity minimum in the center of the pattern is caused by an emerging phase singularity as shown in Fig. \ref{figure1}(d).
Furthermore the azimuthal phase distribution in the vicinity of the central site consists of discrete phase plateaus of $\pi$, which is different to the other phase singularities of the superimposed beam.
For a fixed $a=1$, it is generally possible to introduce an arbitrary phase difference $\phi$, where the defect has a maximal amplitude for $\phi = 0$ and zero amplitude for the case $\phi = \pi$.

\begin{figure}[t]
\centering
\includegraphics[width = \columnwidth]{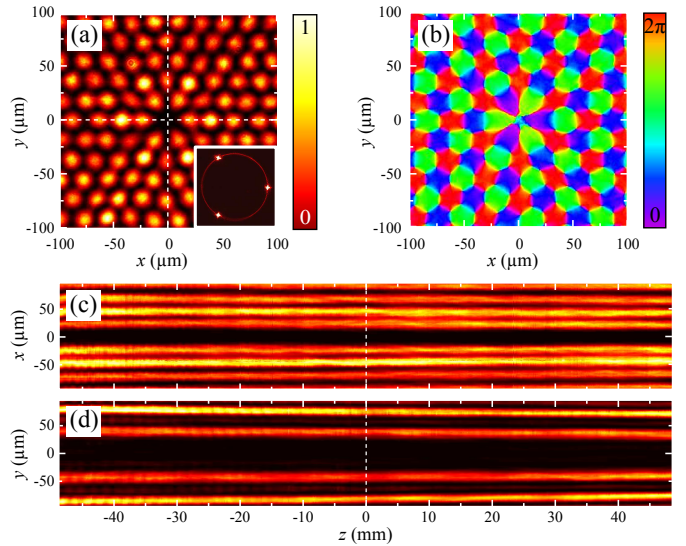}
\caption{(Color online) Experimental realization of a defect beam. (a) and (b): Transverse intensity and phase distribution; dashed lines in (a) mark the positions of the planes shown in (c) and (d), the inset in (a) depicts the intensity distribution of the Fourier spectrum. (c) and (d): Longitudinal intensity distributions of the $x$-$z$ and $y$-$z$ intersection planes; dashed lines mark the longitudinal position of the specified transverse field distribution.
\label{figure2}
}
\end{figure}

In the idealized numerical consideration the ND beam is extended to the entire space, which is not possible in experiment due to the finite dimensions of optical components.
Even though the ND properties of the light field exist in reality only in a specific volume, the range of this volume is large enough for an optical induction in a photosensitive material as can be seen in the 3D examination of the intensity distribution (Fig. \ref{figure2}).
An established and highly flexible method to generate complex light fields of various phase and amplitude distributions is the utilization of a spatial light modulator (SLM).
In our setup similar to the one used in \cite{Boguslawski1}, we project a plane wave emerging from a frequency-doubled Nd:YAG laser with a wavelength of $\unit{532}{\nano\meter}$ onto a phase SLM which modulates the phase as well as the amplitude.
To separate different diffraction orders of the modulated wave an appropriate Fourier-filtering system which is applicable to transmit ND beams with different $k_\bot$ is integrated in the setup.
The determination of the transverse intensity profile for different propagation distances is carried out by a camera mounted on a translation table with a travel distance of $\unit{10}{\centi\meter}$.
In addition the phase of the complex field can be analyzed via superimposing the object wave with a planar reference beam. 
This digital holography technique was for example applied earlier to determine the refractive index change in a photorefractive crystal \cite{Jian-Lin}.
Hereby the first order of the interference pattern recorded by a camera is bandpass filtered and shifted to the center of the Fourier plane.
Finally the subsequent back transformation of the complex field contains the relative phase distribution of the generated beam.
Although the determined phase is available in lower resolution due to the filtering mask in Fourier space, the resulting distribution contains the whole phase landscape including phase vortices.

Figures \ref{figure2}(a) and \ref{figure2}(b) present the transverse intensity and phase distribution of the experimentally implemented superimposed defect lattice, where the lattice constant of the discrete beam was chosen to $g=\unit{25}{\micro\meter}$.
In comparison to the numerical calculated distributions from Figs. \ref{figure1}(c) and \ref{figure1}(d), the experimental results show an excellent agreement in real space as well as in Fourier space.
Furthermore the desired defect site in the center of the pattern reveals the same shape as was predicted in the numerics and displays a fascinating vortex structure exhibiting a multiple topological charge.

To prove the ND properties of the superimposed beam we explored the transverse intensity distribution in different positions with regard to the propagation direction of the light field.
This is realized by shifting the movable camera along an interval of $\unit{10}{\centi\meter}$ with a shift step size of $\unit{1}{\micro\meter}$.
For better visualization of the intensity distribution in dependence of the propagation direction ($z$ axis), we select one particular row and column of each transverse intensity pattern and stack them respectively to the propagation position.
The resulting longitudinal distribution of the intensity in the $x$-$z$ and $y$-$z$ plane are shown in Figs. \ref{figure2}(c) and \ref{figure2}(d), where the dashed lines in the middle depict the transverse intensity and phase distribution of Figs. \ref{figure2}(a) and \ref{figure2}(b).
The constant intensity distribution along the $z$ axis illustrates the ND character of the superimposed lattice with the defect site positioned at the center of the beam.

The intensity profile clearly remains stable. Hence for the optical induction of two-dimensional defect structures in a photorefractive crystal with typical dimensions of several centimeters the presented superimposed lattice beam is highly qualified in view of various experiments.

In conclusion, we have presented a novel type of ND beam with a periodic lattice structure bearing a local defect site by superimposing a periodic light field of hexagonal symmetry with a Bessel beam of zeroth order.
Both the transverse intensity and phase distribution as well as the longitudinal propagation-invariant intensity of the experimentally generated superimposed beam show very good agreement with the simulated field distribution.
Due to the shown properties this defect lattice beam is a perfect tool to parallely induce a photonic defect structure into a photosensitive medium potentially yielding to further investigations of light guidance by defect modes.
The presented work gives only one example of possible superpositions of ND beams.
Due to its high flexibility, the method can be extended to generate more complex defect lattice beams or even other desired nondiffracting field distributions.

\end{document}